\documentclass{emulateapj}

\newcommand{\Ms}{M_\star} \newcommand{\Rs}{R_\star}
\newcommand{\Mbh}{M_{\bullet}} \newcommand{\Mo}{M_{\odot}}

\newcommand{\Rdot}{\mbox{$\dot{R}$}}

\newcommand{\Mdot}{\mbox{$\dot{M}$}}
  \newcommand{\msun}{\mbox{${\rm M}_\odot$}}

\def\apgt{\ {\raise-.5ex\hbox{$\buildrel>\over\sim$}}\ } \def\aplt{\
{\raise-.5ex\hbox{$\buildrel<\over\sim$}}\ }

\shortauthors{HOPMAN, PORTEGIES ZWART \& ALEXANDER}
\shorttitle{ULTRALUMINOUS X-RAY SOURCES}

\bibpunct{\hspace{-4pt}}{}{}{a}{}{}

\begin{document}
\title{Ultraluminous X-ray sources as intermediate mass black holes
fed by tidally captured stars}

\author{Clovis Hopman\altaffilmark{1}, Simon F. Portegies
Zwart\altaffilmark{2, 3} and Tal Alexander\altaffilmark{1}}

\altaffiltext{1}{Faculty of Physics, Weizmann Institute of Science,
POB 26, Rehovot 76100, Israel; clovis.hopman@weizmann.ac.il,
tal.alexander@weizmann.ac.il} \altaffiltext{2}{Astronomical Institute
'Anton Pannekoek', University of Amsterdam, Kruislaan 403, Netherlands}
\altaffiltext{3}{Institute for Computer Science, University of
Amsterdam, Kruislaan 403, Netherlands; spz@science.uva.nl}

\begin{abstract}
The nature of ultraluminous X-ray sources (ULXs) is presently unknown.
A possible explanation is that they are accreting intermediate mass
black holes (IBHs) that are fed by Roche lobe overflow from a tidally
captured stellar companion. We show that a star can circularize around
an IBH without being destroyed by tidal heating (in contrast to the
case of $M_\bullet\!>\!10^6 M_\odot$ massive black holes in galactic
centers, where survival is unlikely). We find that the capture and
circularization rate is $\sim 5\!\times\!10^{-8}\,{\rm yr}^{-1}$,
almost independently of the cluster's relaxation time. We follow the
luminosity evolution of the binary system during the main sequence
Roche lobe overflow phase and show it can maintain ULX-like
luminosities for $>10^7\,{\rm yr}$. In particular, we show that the
ULX in the young cluster MGG-11 in star-burst galaxy M82, which
possibly harbors an IBH, is well explained by this mechanism, and we
predict that $\gtrsim\!10\%$ of similar clusters with IBHs have a
tidally captured circularized star. The cluster can evaporate on a
time-scale shorter than the lifetime of the binary. This raises the
possibility of a ULX that outlives its host cluster, or even lights up
only after the cluster has evaporated, in agreement with observations
of host-less ULXs.

\end{abstract}


\section{Introduction}

Black holes (BHs) have deep potential wells and can transform
gravitational energy very efficiently to radiation.  The energetic
central engines of quasars are thought to host massive black holes
(MBHs) of $M_\bullet>10^6M_\odot$. It is natural to extrapolate this
idea and invoke an intermediate mass black hole (IBH; $10^2\lesssim
M_\bullet/M_\odot\lesssim10^5$) to explain ultraluminous X-ray sources
(ULXs), which are considerably brighter than a stellar mass object
radiating at its Eddington luminosity.  For example, Kaaret et
al. ({\cite{Ka01}) have suggested that an IBH powers the ULX in MGG-11
in star-burst galaxy M82.

The origin of the gas that fuels the X-ray source is unclear, since
almost all the gas in young clusters is rapidly blown away by the
strong winds of massive stars. One possibility for providing the gas
is the tidal disruption of a main sequence (MS) star of mass $\Ms$ and
radius $\Rs$ with periapse $r_p<r_t$, where $r_t = (M_\bullet/\Ms
)^{1/3}\Rs $ is the tidal radius. However, direct disruptions lead to
a short flare ($t_{\rm flare}\lesssim\,\mathrm{yr}$; Rees \cite{RE88};
Ulmer \cite{U99}; Ayal, Livio, \& Piran \cite{ALP00}), which is
incompatible with the $\sim 20\,\mathrm{yr}$ observation period of the
X-ray source. In this {\it Letter} we investigate a more gradual
process for feeding the IBH, namely the tidal capture of a MS star and
the subsequent Roche lobe overflow.

\section{Tidal capture Rate}\label{sec:rate}

A BH in a cluster with velocity dispersion $\sigma$ dominates the
potential within its radius of influence $r_a=GM_\bullet/\sigma^2$;
inside $r_a$ orbits are approximately Keplerian, and stars are
distributed according to a power law $n_\star\propto r^{-\alpha}$,
with $\alpha \approx 3/2$ (Bahcall \& Wolf \cite{BW76}; Baumgardt,
Makino, \& Portegies Zwart \cite{BMPZ03}). The cusp is truncated inside
some radius $r_\mathrm{in}$, e.g. $r_{\rm in}\sim (M_\bullet/\Ms )\Rs
$ where the rate of destructive collisions exceeds the two-body
relaxation rate (Frank \& Rees \cite{FR76}).

Stars can reach an orbit with periapse of order of the tidal radius by
angular momentum diffusion. When the star passes at $r_p$, an energy
$\Delta E_t(r_p)$ is invested in raising tides, causing the star to
spiral in ($r_p\!<\!3r_t$ is typically required for an appreciable
effect). The evolution of a tidally heated star is not
well-understood. Two extreme models of ``squeezars'' (stars that are
continually powered by tidal squeezing) were studied by Alexander \&
Morris (\cite{AM03}). ``Hot squeezars'' are heated only in their outer
layers and radiate their excess energy efficiently; they hardly
expand. ``Cold squeezars'' dissipate the tidal
energy in their bulk and puff up to giant size.

Our analysis is based on the following assumptions.

(1) The stars are ``hot squeezars'' (in \S\ref{sec:disc} we discuss
some consequences of relaxing this assumption).

(2) As long as the eccentricity $e$ is high,
\begin{equation} 
\label{eq:xie}	
1-e=r_p/a<\xi_e\sim0.1,
\end{equation}
where $a$ is the orbital semi-major axis, the stellar structure is not
significantly affected by the tidal heating, and the tidal energy
dissipated per orbit,
\begin{equation} 
\Delta E_t(b)=\frac{G\Ms^2}{\Rs }\frac{T(b)}{b^6}\,,
\end{equation} 
is constant (Alexander \& Morris \cite{AM03}); here $b=r_p/r_t$, and
$T(b)$ is the tidal coupling coefficient, which depends on the stellar
structure and is a strongly decreasing function of $b$ (e.g. Press \&
Teukolsky \cite{PT77}). When the orbit decays to the point where
$1-e>\xi_e$, the tidal heating drops off until eventually the star
circularizes at $a\gtrsim r_t$ and $\Delta E_t\!=\!0$ (Hut
\cite{H80}).

(3) The star can survive as long as its tidal luminosity does not
exceed, to within order unity, its Eddington luminosity $L_E=1.3\times
10^{38}\,\mathrm{erg\,s^{-1}}\Ms /M_\odot$,
\begin{equation} 
\label{eq:xiL}	
\Delta E_t/P< \xi_L L_E ,
\end{equation} 
where $P$ is the orbital period and $\xi_L \approx1$.

The tidal heating rate is highest when $P$ is shortest, just before
tidal heating shuts off when $a=b r_t/\xi_e$
(Eq. \ref{eq:xie}). Therefore, the Eddington luminosity limit
(Eq. \ref{eq:xiL}) corresponds to a minimal periapse
$b_{\mathrm{min}}r_t$ that a star can have and still circularize
without being disrupted, which is given implicitly by
\begin{equation}
\label{eq:bm}
  \Delta E_t(b_{\mathrm{min}}) = \xi_L L_E \frac{2\pi}{\sqrt{G
           M_\bullet}}\left(\frac{b_\mathrm{min}
           r_t}{\xi_e}\right)^{3/2}\,.
\end{equation}
When $r_p<b_{\rm min}r_t$, the star is evaporated by its own tidally
powered luminosity during in-spiral.

Stars within the ``loss-cone'', a region in phase space where stars
have periapse smaller than $r_t$ (Frank \& Rees \cite{FR76}; Lightman
\& Shapiro \cite{LS77}), are disrupted by the BH. Two-body scattering
sustains a flow of stars in angular momentum space towards the
loss-cone. During in-spiral, two-body interactions change the periapse
of the star. The time $t_p$ over which the periapse of a star is
changed by order unity due to many small angle deflections is
(Alexander \& Hopman \cite{AH03})
\begin{equation}
\label{eq:tp}
t_p(b, a)= \frac{br_t}{a} t_r\,,
\end{equation}
where $t_r$ is the relaxation time. Note that $t_r$ does not depend on the distance from the BH for $\alpha=3/2$.

The in-spiral time is the time it takes until the semi-major axis of the star becomes formally zero; for a hot squeezar it is (Alexander \& Hopman
\cite{AH03})
\begin{equation}
  t_0(b, a) = \frac{2\pi \Ms \sqrt{G M_\bullet a}}{\Delta E_t(b) }\,.
\end{equation}
If deflections increase the periapse, the dissipation becomes much
less efficient, while if the periapse decreases, the star may cross
$r_t$ and be disrupted. Either way it fails to
circularize. Circularization can happen only if the in-spiral time
$t_0$ is shorter than the time-scale for deflections, $t_p$. The
widest orbit $a_c(b)$ from which a star can still spiral in for
periapse $b r_t$ is given by $t_0(b, a_c)=3t_p(b, a_c)$ (Alexander \&
Hopman \cite{AH03}, Eq. [11] for $\alpha = 3/2$). It then follows from
Eq. (\ref{eq:bm}) that the maximal distance $a_{\mathrm{max}}$ from
which a star can originate to reach the tidal radius without being
destroyed is
\begin{equation}\label{eq:amax}
    a_{\mathrm{max}} = \left[\frac{3\Delta
	E_t(b_\mathrm{min})b_\mathrm{min}r_tt_r}{2\pi \Ms \sqrt{G
	M_\bullet}}\right]^{2/3}.
\end{equation}
Within $r_\mathrm{in}$ the cusp flattens and relaxation is
inefficient, so there are hardly any stars on eccentric orbits.  Since
$r_{\rm in}$ grows more rapidly with $M_\bullet$ than $a_{\rm max}$,
there exists a maximal BH mass $M_\mathrm{max}$, such that for
$M_\bullet>M_\mathrm{max}$, $a_{\mathrm{max}}<r_\mathrm{in}$, and
tidal capture is strongly suppressed. Fig. (\ref{fig:radii}) shows
$a_\mathrm{max}$ and $r_\mathrm{in}$ as a function of $M_\bullet$; for
the calculation of $t_r$ we assumed that the $M_\bullet - \sigma$
relation
\begin{equation}\label{eq:Msigma} 
  M_\bullet = 1.3\times10^8 M_\odot\left
      ( \frac{\sigma}{200\,\mathrm{km\,s^{-1}}}\right)^4
\end{equation}
(Ferrarese \& Merritt \cite{FM00}; Gebhardt et al. \cite{Gea00};
Tremaine et al. \cite{Tr02}) can be extended to IBHs (see e.g.,
Portegies Zwart \& McMillan \cite{PZMcM02}). Circularization is only
possible for $M_\bullet<M_{\rm max}\approx10^5M_\odot$.

\begin{figure}
\plotone{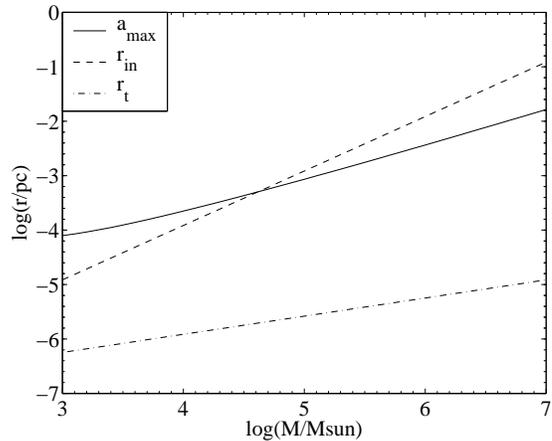} 
\caption{Dependence of $a_\mathrm{max}$ (solid), $r_\mathrm{in}$ (dashed), and $r_t$
(dashed-dotted) on the mass of the BH, for a $10\Mo$ star. Circularization is only
possible provided that
$a_\mathrm{max}>r_\mathrm{in}$. \label{fig:radii}}
\end{figure}

The rate $\Gamma$ at which stars diffuse into orbits that allow
successful circularization is given by (Eq. [9] Syer \& Ulmer
\cite{SU99})
\begin{equation}\label{eq:rate}
  \Gamma = \frac{(a_\mathrm{max}/r_a)^{3 - \alpha}N_a}
	        {t_r\mathrm{ln}(2\sqrt{a_\mathrm{max}/b_\mathrm{min}r_t})}
	        \,\,\,\,\,\,\,\,\,\,\,\,(a_{\mathrm{max}}>r_\mathrm{in}),
\end{equation}
where the logarithmic term expresses the depletion of the stellar
density near the loss-cone; $N_a$ is the number of stars within the
radius of influence. The rate is essentially independent on $\Ms$ for a fixed stellar mass within $r_a$, and it decreases only logarithmically
with $t_r$ (cf. Eqs [\ref{eq:amax}, \ref{eq:rate}]): a larger $t_r$
increases the volume of stars that contributes to $\Gamma$, but
decreases the rate at which stars enter the loss-cone. The rate does
not depend very sensitively on our assumptions: roughly,
$\Gamma\propto\xi_L/\xi_e$.

\section{Roche lobe overflow on the main sequence}\label{sec:RLOF}
Orbital angular momentum conservation implies that the circularization
radius is $a_{\rm circ}\!=\!2 b_{\rm min} r_t$. Efficient in-spiral
and successful circularization require $b_{\rm min}\!\sim\!2-2.5$, so
that $a_{\rm circ}\!\sim\!(4-5)r_t$. The onset of mass transfer
through the Roche lobe occurs when the distance between the IBH and
the star is $a_{\rm circ}\!\sim\!2 r_t$ (assuming $\Ms\!=\!10M_\odot$,
$\Ms/\Mbh\sim0.01$, Eggleton \cite{Eg83}). This is roughly a factor of
two smaller than the typical value of $a_{\rm circ}$. However, as it
evolves on the MS, a $10M_\odot$ star expands by a factor of
$\sim\!2.7$ by the time it reaches the terminal age MS (TAMS). This
implies that Roche lobe overflow (RLOF) occurs at some point on
the MS, and continues for a time $t_{\rm x}$, which is shorter than
the MS lifetime $t_{\rm MS}$ (we assume here that the star expands
significantly only after it has circularized). In the following
analysis we assume for simplicity that RLOF holds over the entire
MS. This does not affect our conclusions significantly since the
observationally relevant phase of high X-ray luminosity occurs toward
the TAMS. For massive MS stars RLOF is preceded by a less 
luminous phase resulting from the accretion of strong stellar winds.

Mass transfer from a MS star to an IBH is driven by the thermal
expansion of the donor and the loss of angular momentum from the
binary system.  Mass transfer then implies that the donor fills its
Roche lobe ($R_{\star} = R_{\rm Rl}$) and continues to do so
($\Rdot_{\star} = \Rdot_{\rm Rl}$). We assume that as long as the
Eddington limit is not exceeded, all the mass lost from the donor via
the Roche lobe is accreted by the IBH ($\Mdot_{\bullet} =
-\Mdot_{\star}$). Otherwise, the mass in excess of the Eddington limit
is lost from the binary system.

The expansion of the donor on the MS is calculated using fits from
Eggleton, Tout \& Fitchett (\cite{ETF89}) to detailed stellar
evolution calculations. We assume that the evolution of the donor
was not affected by mass loss.  Variations in the Roche radius of the
donor can be computed from the redistribution of mass and angular
momentum in the binary system. The radius of the Roche-lobe is
estimated with the fitting formula from Eggleton (\cite{Eg83}).

We stop following the binary evolution at the TAMS; the simple model
for calculating the amount of mass transfer may be inappropriate for
the post-MS evolution of the donor, as the star then rapidly expands.
However, at the end of the MS the donor still has a considerable
envelope and the star ascends the giant branch. The post-MS
evolution is likely to result in a short ($t_\mathrm{PMS}<0.1\, t_{\rm
MS}$) phase in which the luminosity increases by more than an order of magnitude.

In Fig. (\ref{fig:Mdon}) we plot the mass of the donor as a function
of time. It is assumed here that there is Roche lobe contact directly
after circularization. As discussed, this actually only happens
when the star has evolved towards a later stage of the MS
phase, which is when the mass loss in the plot starts to drop more
rapidly.

\begin{figure}
\includegraphics[angle=-90, width =
0.45\textwidth]{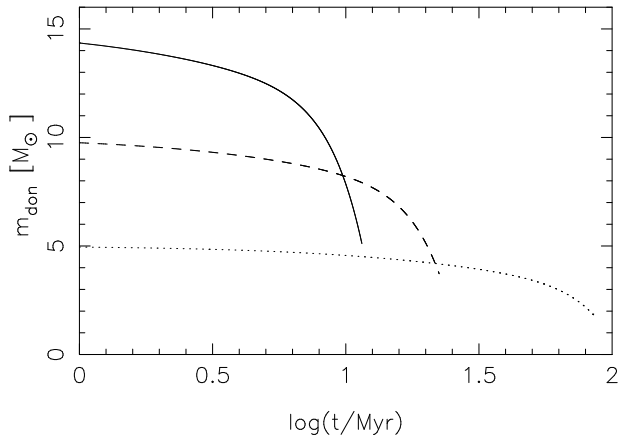}
\caption{Mass evolution of the donor star, assuming Roche-lobe
contact at the zero-age MS to a $1400 M_\odot$ IBH.  The solid, dashed
and dotted curve are for a $15 M_\odot$, $10 M_\odot$, and $5 M_\odot$
donor.\label{fig:Mdon}}
\end{figure}

We estimate the X-ray luminosity during mass transfer with the model
discussed by K\"ording, Falcke, \& Markoff (\cite{KFM02}). They argue
that the X-ray luminosity is generated by an accretion disk. The
binary is in the hard state if $\Mdot> \Mdot_{\rm crit}$, in which
case $L_{\rm x} = \epsilon \Mdot c^2$. At lower accretion rates
$L_{\rm x} = \epsilon \Mdot c^2 \Mdot/\Mdot_{\rm crit}$, in which case
the X-ray source becomes transient (i.e. short outbursts, separated by long states of quiescence, Kalogera et al.\, 2003).  For
$\Mdot_{\rm crit}$ we adopt the equation derived by Dubus et al.\,
(\cite{DU99}, see Eq. 32) and assume $\epsilon = 0.1$.  These choices are
comparable to $\Mdot_{\rm crit} \sim 10^{-7} M_\odot\, \mathrm{yr}^{-1}$ of K\"ording
et al.\, (\cite{KFM02}).  The resulting X-ray luminosity is presented
in Fig.\,(\ref{fig:Lx}).  Note that lower mass donor binaries
($\Ms \aplt 5$\,\msun) live longer, are less luminous and tend to show transient behavior, where
high mass donor binaries are more luminous, shorter lived, steady sources.

\begin{figure}
\includegraphics[angle=-90, width =
0.45\textwidth]{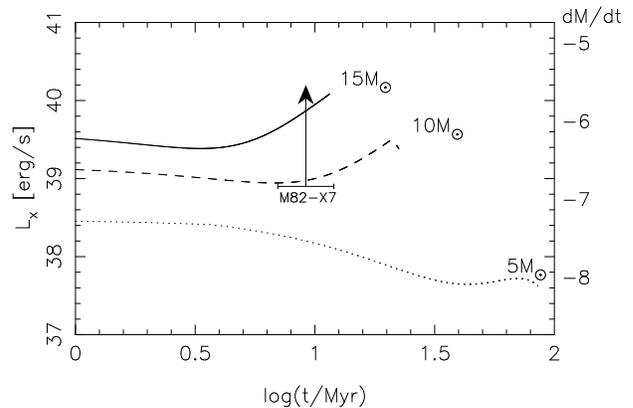}
\caption{X-ray
luminosity for a $1400 M_\odot$ accreting IBH as a function of time.
Line styles as in Fig.\,\ref{fig:Mdon}.  To the right side of the
figure we added $\dot{M}$ in logarithmic units of
$M_\odot\,{\rm yr^{-1}}$ for the regime where $\Mdot > \Mdot_{\rm
crit}$. The radius of the star grows significantly towards the TAMS
(where the lines for 15\msun\, and 10\,\msun\, donors rise). It is probably only near that point at which
RLOF actually starts, so that the luminosity is higher than would be
estimated from RLOF of a zero age MS star.  The 5\,\msun\, donor does
not show this rise in $L_{\rm x}$ as $\dot{M}<\Mdot_{\rm crit}$ after $\sim 10$\,Myr. This also causes the  drop of $L_{\rm x}$ for the 10\msun\, donor star at the end of its evolution. When  $\dot{M}<\Mdot_{\rm crit}$, the source becomes transient.
  \label{fig:Lx} }

\end{figure}
\section{Cluster MGG-11}\label{sec:MGG-11} 
We apply our analysis to the young dense star cluster MGG-11 in the
irregular galaxy M82, at a distance of $\sim 4$ Mpc. This cluster
contains the variable X-ray source M82-X7 with $L_{\rm x} = (0.8-160)
\times 10^{39}\, {\rm erg s^{-1}}$ (Watson, Stanger \& Griffiths
\cite{WSG84}; Matsumoto \& Tsuru \cite{MT99}; Kaaret et
al. {\cite{Ka01}). The velocity dispersion in the cluster is
accurately measured, $\sigma=11.4\pm0.8 \,\mathrm{km\, s^{-1}}$
(McCrady, Gilbert, \& Graham \cite{MGG03}). We assume that this cluster contains an
IBH which is also the engine for the X-ray source. If this IBH obeys
the $M_\bullet - \sigma$ relation (\ref{eq:Msigma}), its mass is
$M_\bullet = 1.4\times10^3M_\odot$, which is consistent with the
recent calculations of Portegies Zwart et al. (\cite{PZ04}), who show that an IBH could have formed dynamically in MGG-11 by a runaway merger of MS stars. Within
its radius of influence $r_a=0.05 \,\mathrm{pc}$, the number of stars
is $N_a=2M_\bullet/\Ms $ (Merritt \cite{Me03}) and $t_r\sim10^5-10^6\,\mathrm{yr}$.

The age of the cluster is $t_{cl}= (7-12)\,\mathrm{Myr}$,
corresponding to a turn-off mass of $17-25 M_\odot$ (Eggleton et
al. \cite{ETF89}). The mean stellar mass of the cluster at birth
is $\langle M_\star \rangle=3\,M_\odot$, but as a result of
mass-segregation the average mass within $r_a$ is much higher. The
direct N-body calculations of Portegies Zwart et al. (\cite{PZ04}) show that at an age of 7
Myr, the mean mass of the single stars in the core of MGG-11 is
$\langle M_\star \rangle = 8\pm3 M_\odot$. For simplicity we assume
within $r_a$ a single mass population of stars with $\Ms =10M_\odot$,
and radius $\Rs = 5.4 R_\odot$ (Gorda \& Svechnikov \cite{GS98}).  The
results of Fig. (\ref{fig:Lx}) show that a MS donor of mass $\Ms
\gtrsim10\,M_\odot$ can account for the luminosity of the ULX in
MGG-11.

We assume $(\xi_e, \xi_L)=(0.1, 0.5)$, and take the numerical values
for the function $T(b)$ for parabolic orbits from Alexander \& Kumar
(\cite{AK01}).  With these parameters we find a capture rate of
$\Gamma=5\times10^{-8}\, \mathrm{yr^{-1}}$. This implies that a
fraction $\Gamma\,t_{cl}t_{\rm x}/t_{\rm MS}= (30-50)\%$ of clusters
harboring an IBH has formed a tidal binary and may be observed during
RLOF in the MS phase. A fraction $\Gamma t_{cl}t_{\rm PMS}/t_{\rm
MS}\sim 4 \%$ of clusters with an IBH should be observed during the
more luminous post-MS phase.

\section{Summary and discussion}
\label{sec:disc}

MS stars can spiral into an IBH as a result of tidal capture and
circularize close to the tidal radius. This process is unique to IBHs,
since stars cannot survive tidal in-spiral around a MBH in a galactic
center. After circularization, the star expands on the MS until
high luminosity RLOF accretion starts toward the end of the MS
phase. We analyzed RLOF during the MS phase in some detail and
calculated the X-ray luminosity. Post-MS RLOF is harder to model, but
the resulting luminosity is expected to be at least an order of
magnitude brighter and about an order of magnitude shorter in
duration. The X-ray luminosity is consistent with observed ULXs, such
as MGG-11.

MGG-11 is the only cluster out of hundreds in M82 with a ULX. Possibly
other clusters were not sufficiently dense to form IBHs (Matsushita et
al. \cite{Mea00}; Portegies Zwart et al. \cite{PZ04}). If a fraction
$f_\bullet$ of the $N_{cl}$ clusters in M82 harbors an IBH, the number
of ULXs is estimated by $N_{\rm x} = f_\bullet \Gamma t_{cl}t_{\rm
x}/t_{\rm MS}$. Thus $f_\bullet$ has to be of the order of a few
percent in order to account for one ULX in M82.

In order to circularize, a star has to dissipate $\sim(M_\bullet/\Ms
)^{2/3}$ times its binding energy. If a certain fraction
$\delta$ of the energy is invested in bulk heating (for ``hot
squeezars'' $\delta=0$ as assumed so far, for ``cold squeezars'' $\delta=1$), the star expands. An X-ray binary can form only if $\delta\!<\!(\Ms
/M_\bullet)^{2/3}$. Nevertheless, a shorter lived ULX-phase is
still possible even if $\delta> (\Ms /M_\bullet)^{2/3}$. When the star
expands to a radius $>\!b\Rs$ it is gradually peeled every
periapse passage and feeds the IBH for a period much longer than
$t_{\rm flare}$. However, the process is limited by the two-body
deflection time-scale $t_p$, which is $\lesssim\!10^3\,\mathrm{yr}$
for a $10^3M_\odot$ IBH. This translates to a detection probability of
only $\Gamma t_p\!\sim\!5\!\times\!10^{-5}$, and so it is unlikely
that the ULX in MGG-11 originates in this type of process. For similar
reasons, it is improbable to observe a very luminous tidally heated
star (squeezar) during the final stages of its in-spiral into an IBH
in a stellar cluster (this may be possible for squeezars near the MBH
in the Galactic Center, where the in-spiral time is longer and the
capture rate higher, Alexander \& Morris \cite{AM03}).

The lifetime of the host cluster is limited by the galactic tidal
field, and can be as short as 100\,Myr (Portegies Zwart et
al. \cite{PZ01}). This is much shorter than the RLOF phase of a low mass
donor (e.g. $\sim$ Gyr for a $2 M_\odot$ star). Thus, the X-ray
life-time of a low-luminosity binary can be much longer than the
life-time of the cluster. Quiescent orphaned IBHs can suddenly light
up when their companion ascends the giant branch and starts to
transfer mass to the IBH. Our scenario predicts the existence of
host-less ULXs, which are more likely to be transient and less luminous. Their exact fraction in the ULX population cannot be
reliably estimated at this time. However, it is interesting to note
that $3-10$ out of $14$ of the ULXs in the Antennae Galaxies are
coincident with a stellar cluster, while the others are not (Zezas et
al. \cite{ZFRM02}).

\acknowledgements
We thank the referee, Monica Colpi, for comments that improved the manuscript.
This work is supported by the Royal Netherlands Academy of Sciences
(KNAW), the Dutch organization of Science (NWO), the Netherlands
Research School for Astronomy (NOVA), ISF grant 295/02-1, Minerva
grant 8484, and a New Faculty grant by Sir H. Djangoly, CBE, of
London, UK.

\end{document}